# High diffusivity pathways govern massively enhanced oxidation during tribological sliding

Julia S. Rau[a,b], Shanoob Balachandran[c], Reinhard Schneider[d], Peter Gumbsch[a,e,] Baptiste Gault[c,f], Christian Greiner[a,b*]

[a] Institute for Applied Materials (IAM), Karlsruhe Institute of Technology (KIT), Kaiserstrasse 12, 76131 Karlsruhe, Germany; Julia.rau@kit.edu

[b] IAM-CMS MicroTribology Center µTC, Strasse am Forum 5, 76131 Karlsruhe, Germany

[c] Max-Planck-Institut für Eisenforschung GmbH, Department of Microstructure Physics and Alloy Design, Max-Planck-Straße 1, 40237 Düsseldorf, Germany, b.gault@mpie.de

[d] Laboratory for Electron Microscopy (LEM), Karlsruhe Institute of Technology (KIT), Engesserstrasse 7, 76131, Karlsruhe, Germany, Reinhard.schneider@kit.edu

[e] Fraunhofer IWM, Woehlerstrasse 11, 79108 Freiburg, Germany, Peter.gumbsch@kit.edu

[f] Department of Materials, Royal School of Mines, Imperial College London, Prince Consort Road, London, SW7 2BP, UK

*Corresponding author: greiner@kit.edu, +49 72120432742

**Graphical Abstract:**

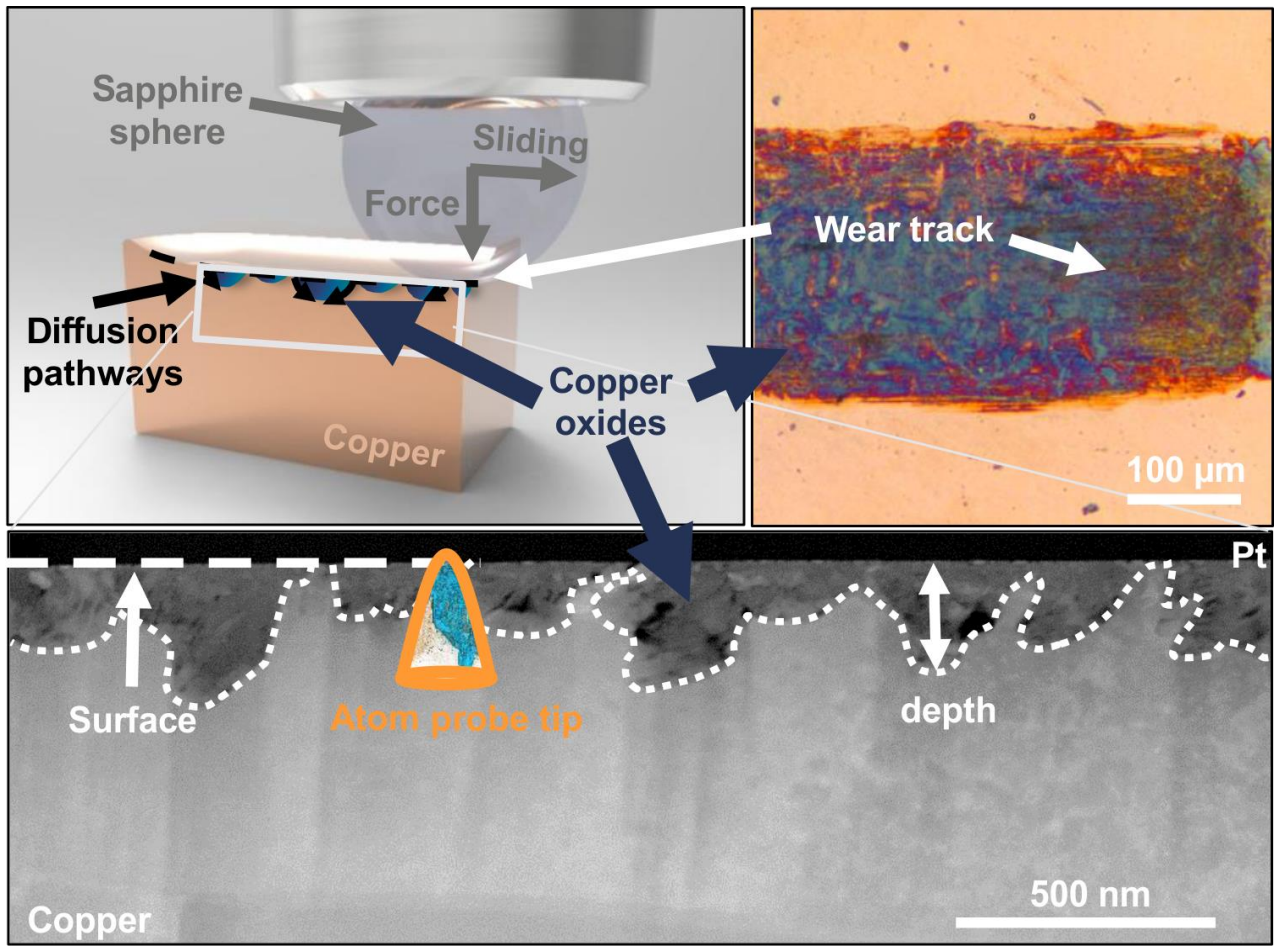

**Abstract:**

The lifetime of moving metallic components is often limited by accelerated oxidation. Yet, the mechanisms and pathways for oxidation during tribological loading are not well understood. Using copper as a model system, tribologically-induced oxidation is systematically investigated by varying the sliding speed and test duration under mild tribological loading. We demonstrate that tribo-oxidation is controlled by test duration rather than the number of cycles or the sliding speed. Plastic deformation from tribological loading creates dislocations, grain and phase boundaries that act as high diffusivity pathways. A combination of electron microscopy and atom probe tomography revealed significantly enhanced atomic concentration of the diffusing species around dislocations. Oxygen diffusion into the bulk as well as of copper towards the free surface along these defects control the oxide formation kinetics. Our work paves the way for formulating a physics-based understanding for tribo-oxidation, which is crucial to develop strategies to slow or decrease oxidation and to strategically tailor surfaces to increase the lifetime of engineering systems.

# 1. Introduction:

Around 20% of the world's primary energy production is needed to overcome friction forces[1] and ~ 2 - 4% of the GDP are lost to corrosion in industrialized countries[2]. Tribologically-induced oxidation often is responsible for undesirable friction and wear properties[3] and component failure[4,5]. Owing to its significance, tribo-oxidation has long been investigated and it is clear since the 1960ies that the growth rate of the oxide film is significantly faster than under static conditions[6]. Since diffusion processes must be involved in the oxide growth, the enhanced oxide formation under tribological conditions has mostly been attributed to a sliding-induced temperature increase[7,8]. Alternatively, oxidation has been related to a change in contact-duration between the sliding surfaces, resulting in the formation and growth of micro welds[9,10] as well as the time available for out-of-contact oxidation in oxidative wear[11]. Over all, it remains elusive why and how oxides grow so quickly under tribological loading. Efforts to unveil the fundamental underpinnings of the tribo-oxidation, have focused on studies of model contacts, for example between high-purity copper and sapphire, where first oxygen-rich cuprous-oxide ($Cu_2O$) hemispherical features were found below the surface, that later grow and coalesce into a continuous oxide layer on the surface[12]. Under the employed mild loading conditions, a significant temperature increase in the contact can safely be excluded, which sharpens the question for the mechanistic origin of the enhanced oxide growth.

Our aim is to identify the relevant mechanisms for this enhanced growth. To go beyond previous analyses, we performed a systematic set of model experiments on the copper-sapphire system with the setup shown in **Figure 1** and deployed near-atomic-scale microscopy and microanalysis to reveal the detailed physico-chemical mechanisms underpinning tribologically-activated acceleration of the oxidation process.

## 2. Experimental:

**Materials.** Since copper is a well investigated model system to understand the formation of metals oxides, oxygen-free high conductivity (OFHC) copper samples with a purity > 99.95% (Goodfellow, Friedberg, Germany, 25 x 6 x 5 mm³) were investigated. The average grain size of the polycrystalline samples was 30 – 40 µm. Additionally, a copper single crystal (Mateck, Jülich, Germany) with a purity of 99.999% was investigated. A ±<0-11> sliding direction was chosen on a (111) surface since this surface shows the slowest native oxidation[13]. Sapphire spheres were used as counter bodies (diameter of 10 mm, Saphirwerk, Bruegg, Switzerland) for their high hardness and chemical inertness.

**Sample preparation.** Sample preparation including heat treatment, grinding, polishing and electropolishing are described in detail in[12,14]. The final electro-polishing was performed right before the tribological tests to achieve a sample surface with minimal native oxidation from the environment.

**Tribological testing.** The setup of our reciprocating linear tribometer (**Figure 1**a) is described in [12,14,15]. A normal load of 1.5 N was applied on the sapphire sphere by a dead weight. The copper samples were moved by a linear motor (M-403.2PD, Precision Translation Stage, Physik Instrumente, Karlsruhe, Germany). Experiments were performed with sliding distances of 2,4 mm to 12 mm in a reciprocating fashion for 1,000 to 5,000 cycles. The tests were conducted in air at a relative humidity of 50 %, at room temperature and ambient pressure without lubrication. Humidity and temperature were controlled during the tribological experiment. The sliding speed was varied from 0.1 mm/s to 5.0 mm/s. One additional experiment was performed at a higher speed of with 50 mm/s, for which a high speed stage (PIMag® High-Load Linear Stage, V-412.136025E0, Physik Instrumente, Karlsruhe, Germany) was used. Experiments run with a sliding speed of 1.5 mm/s were exposed to the ambient air after the tribological loading for 2.5h, 24h and 48h to investigate the influence of exposure time. The exposure time represents the time for (further) oxidation in a controlled environment after the tribological loading occurred. One experiment was run in $N_2$ and exposed afterwards 48h to air with 50% relative humidity. Besides the samples for the exposure time experiments, all samples were stored in a desiccator constantly being evacuated to a pressure below 1 mbar immediately after the end of each experiment.

**Microstructure characterization**. After 1,000 cycles, the surfaces were analyzed to systematically investigate the resulting tribo-oxides. Cross-sections and transmission electron microscopy (TEM) foils were prepared right after the tribological tests or after a defined exposure time using a focused ion beam / scanning electron microscope (FIB/SEM, FEI Helios 650, ThermoFisher Scientific, Waltham, Massachusetts, USA). The cross-sections were cut along or perpendicular to the sliding direction and perpendicular to the sliding surface in the center of the wear tracks (see SI of [12]). In order to protect the sample surface from ion beam damage when milling, two protective platinum layers (one with the electron and one with the ion beam) were deposited. The boundary between the platinum layer and the sample surface is marked by a dashed line (**Figure 1,2,6**). In order to characterize the oxides further, the oxide thickness is measured as indicated by the white arrow in **Figure 1**c. Each data point represents one experiment where one cross-section was investigated. On each cross-section, five to eleven images were evaluated and the mean values of the maximum oxide thicknesses *d* were determined.

**Chemical composition characterization.** Energy-dispersive X-ray spectroscopy (EDS) was performed inside a 200 kV transmission electron microscope (TEM) (FEI Tecnai Osiris equipped with a ChemiSTEM system) with four silicon drift detectors (Bruker) in the scanning TEM (STEM) mode for detailed microchemical analysis. The measurement period was 10 – 25 minutes, while a possible drift was automatically corrected by a cross-correlation with reference images. The maximum lateral resolution in the EDS measurements was approximately 1 nm. Two TEM foils for two experiments with different exposure times were analyzed (1.5 mm/s, 1,000 cycles, 0h and 48h). A single crystal run for 5,000 cycles with a sliding speed of 0.5 mm/s was analyzed by atom probe tomography (APT) (Cameca Local Electrode Atom Probe 5000XR). A protective Cr-layer was deposited on the wear track before APT sample preparation. The Cr-layer is not shown in the APT results for clarity. The acquisition was done in laser mode with a specimen temperature of 60 K. Further parameters were: pulse rate 200 kHz, pulse energy 40 pJ, detection rate 0.5 - 1%, detection efficiency ~ 57%.

## 3. Results and Discussion:

### 3.1 Oxide formation under tribological load

The experimental set-up used in all experiments is depicted schematically in **Figure 1**a. A coloring of the wear track after sliding clearly reveals oxide formation under loading in contrast to the unloaded copper-colored surface next to the wear track (**Figure 1**b). The rate of oxide growth under tribological loading can be several hundred times faster than copper's native oxidation under the same environmental conditions (SI1). The influence of the sliding speed, test duration and exposure time on oxide growth is quantified via the thickness to which oxides are found below the surface for 1000 cycle tests (**Figure 2 and Figure 3**). We independently varied the sliding speed (0.1, 0.25, 0.5, 1.5, 2.5, 5.0 and 50.0 mm/s) and the sliding distances, which for a sliding speed of 0.1 mm/s led to different test durations from 13.5 to 67 hours. Finally, the exposure time to the controlled environment was varied from 0 to 48 hours (0, 2.5, 24 and 48 hours). While previous XPS analysis of a sapphire counter body reported no copper wear particles on the spheres after 1,000 sliding cycles with 0.5 mm/s[16], our experiments showed some wear particles on the sapphire sphere after sliding (**Figure 4**). Both, after 1,000 and 5,000 sliding cycles, small particles around the contact area were visible which is possibly due to the interaction of copper oxides with the sapphire sphere[17]. Previous EDS investigations did not show any aluminum inside the wear track[14,17].

**Figure 2**a,b and **Figure 3**a show a significantly increased oxide thickness at lower sliding speeds. For speeds below 0.5 mm/s, the oxide thickness correlates with the time for a single trace (solid line in **Figure 3**a) and then deviates for higher speeds. To decouple the effects of sliding speed and test duration, we reduced the sliding distance (from 12.0 mm to 9.9, 7.2, and 2.4 mm) for a constant sliding speed of 0.1 mm/s (**Figure 2**c,d and **Figure 3**b blue data points). The oxide thickness increased linearly with the test duration (**Figure 3**b). Above 60 hours, crack formation within the oxide and wear particle formation was observed (SI2). Samples subjected to 1,000 cycles at 1.5 mm/s and left in ambient environment for up to 48h showed no significant ongoing oxide growth without further tribological loading (**Figure 2**e,f, **Figure 3**c). Hence, the oxide thickness increases due to the tribological load, but not or only very little with a further exposure to the ambient environment.

Oxidation and tribo-oxidation processes typically follow a growth law where the rate of oxidation decreases with oxidation time (parabolic rate law)[18]. Astonishingly, a quasi-linear behavior with test duration is observed for tribo-oxidation here (**Figure 3**b), followed by an almost constant oxide thickness after unloading and exposure to air (**Figure 3**c). The question arises which mechanism is behind this unexpected observation. The strong dependence on time at constant number of cycles, excludes a purely mechanical origin of the oxide growth, like folding (e.g. [19]), and rather suggests that diffusion processes play a major role in oxide formation, especially at sliding speeds below 0.5 mm/s. In this regime, any excess heat, possibly generated in the contact certainly did not lead to a significant temperature increase but was transferred away due to the excellent thermal conductivity of copper (SI3). For low temperature oxidation (< 180 °C) of copper single crystals, Fujita et al. found that activation energies of oxidation agreed with the activation energies of diffusion (along surface and grain boundaries)[20]. They concluded that low-temperature oxidation kinetics can be attributed to surface and grain boundary diffusion on and within the oxide[20]. As a first approximation, we thus consider our results as predominantly diffusion controlled. Phase and grain boundaries, dislocations and the lattice may act as diffusion pathways.

Driving forces for diffusion may be the concentration gradient between the surrounding atmosphere and the high-purity copper, compositional variations within the formed oxides[12] or induced strains through plastic deformation. The resulting structure under and after sliding is certainly far from thermodynamic equilibrium. Oxygen and copper ions therefore experience significant driving forces to diffuse inside the tribologically-deformed subsurface and the oxides which were formed during the tribological loading. For example, previous electron microscopy analyses revealed nanocrystalline $Cu_2O$ in an amorphous matrix[12]. Compositional variations in-between (1) the surface and the copper bulk as well as between (2) the nanocrystalline areas and the amorphous parts, imply differences in the chemical potential that drive these diffusional processes and the associated changes in size and morphology of the oxide layer.

As the most simple approach, one can use the Einstein relation $D \sim \frac{d^2}{2t}$, where $t$ is the test duration and $d$ the oxide thickness[21], to compare the diffusion coefficient $D_O^{Cu}$ of oxygen in copper for tribo-oxidation and native oxidation. This approach yields a four to five orders of magnitude increase for the effective diffusion coefficients for tribo-oxidation compared to the values for native-oxidation (see details in SI4). For room temperature, this is consistent with reports of self-diffusion along dislocations, the carriers of plastic deformation, in metals and metal oxides[22,23] (see SI5). It therefore appears reasonable to assume that plastic deformation due to tribological loading creates high diffusivity pathways. Diffusion through grain boundaries, phase boundaries and dislocations all have a lower activation energy than bulk diffusion through the lattice and have been recognized as important elements for diffusion during the plastic deformation of copper[24,25]. With the connection to the plastic deformation, pipe diffusion along dislocation cores could be the key mechanism leading to accelerated oxygen penetration into copper and resulting in an enhanced oxidation kinetics.

### 3.2 Diffusion pathways

Grain and phase boundaries, dislocations or the lattice itself act as diffusion pathways. To identify the dominating diffusion pathway when oxides are present, we consider the following (**Figure 5** and **Table 1**): With Fick's first law[26]

$$j_x = -D_x \frac{\Delta c}{\Delta x} = \frac{1}{A_x} \frac{\Delta N_x}{\Delta t}$$

the number of diffusing atoms per second $N_x$ for each pathway can be estimated ($j_x$ = diffusive flux, $D_x$ = diffusion coefficient). In this example, the diffusing species is set as oxygen atoms for simplicity. $X$ represents lattice, phase boundary (here grain boundary) and dislocation pathways. $A_x$ is the area of the respective pathway. The grain boundary width is set to $t = 0.6$ nm in accordance with literature[27]. With $\Delta c = \frac{n}{V} = \frac{4 * 0.95}{(0.361\ nm)^3}$ and $\Delta x = -50$ nm (after 1,000 sliding cycles), the resulting amount of atoms through each pathway is calculated and shown in **Table 1**. The oxygen concentration at the surface is estimated as 100 at% and 5 at% (between 3 – 5 at% in TEM-EDS and APT results, **Figure 6**c+f, **Figure 7** and **8**) in the bulk below the oxide. **Table 1** compares the amount of diffusing atoms per second for the in **Figure 5** shown oxide. Three different cases are considered: through the lattice, through the phase boundary only and through one single pipe. No leakage models are considered here[28]. These data demonstrate that a significant number of atoms can enter the material through dislocation pipe-diffusion ($1 \cdot 10^4$ atoms/s) and phase boundary diffusion ($3 \cdot 10^5$ atoms/s), while lattice diffusion ($5 \cdot 10^{-16}$ atoms/s) can be neglected. Inserting the approximate oxide particle size from **Figure 5**, the transport along the phase boundary and the transport along dislocations are equally important if about 20 dislocation pipes are inside the oxide. This corresponds to a dislocation density of ~ $7 \cdot 10^{10}$ 1/cm$^2$. Such a density is in very good agreement with a dislocation density in deformed single and polycrystalline copper[29],[30]. Liu et al.[16] observed geometrically necessary dislocation densities in copper after 1 pass of a sapphire sphere (2N) of $4 \cdot 10^{10}$ to $10^{12}$ 1/cm². For higher densities than $10^{10}$ 1/cm², dislocations become the dominating diffusion pathway. Despite our basic assumptions (stationary diffusion flux, no concentration change considered), the calculations show the relevance of dislocations as diffusion pathways inside the oxides. The previous example shows: While lattice diffusion is negligible for the examined temperature and test duration, pipe diffusion along dislocation cores can contribute to the penetration depth of the diffusing oxygen at least as much as phase boundary diffusion. Diffusion along defects can be described by the models of Fisher and Smoluchowski for grain boundaries and dislocations[31,32]. However, the diffusion length along a defect decreases with increasing time [28] which would, for long oxidation times, not result in the observed quasi-linear oxide growth with increasing test duration but a parabolic behavior (**Figure 3**b). A linear oxide growth law is reported for very thin or porous oxide layers, where diffusion processes are not rate-determining but the reaction of oxygen with the metal at the metal/air phase boundary[33]. Both scenarios (thin or very porous oxide) do not apply for all cycles numbers. Hence, the quasi-linear oxide growth with increasing test duration is believed to have its origin in an anomalous (non-parabolic) diffusion behavior. Diffusion processes with $\Delta x^2 \sim t^\alpha$ ( $x$ = diffusion distance) with α > 1 are called 'super diffusion' and for α = 2 ballistic diffusion[34]. For example, ballistic transport of charge carriers is reported along dislocations in Si[35]. For impurity elements, one possible explanation for super diffusion processes may be the diffusion with 'traps' within defects: Defects may serve as traps for the diffusing species due to the deepening of the associated potential well at specific sites[36–38]. Trapping of a diffusing species was intensely studied for hydrogen in steels[38,39], however it is not limited to this system but also valid for oxygen[40].

With unfilled traps, the absorption kinetics of a species is significantly faster and associated with a higher apparent diffusivity than the evolution kinetics (inequality of fluxes)[39]. With increasing trap saturation, blocking and correlation effects may decrease the diffusivity again [41,42]. Finally, as a result of trapping for large concentrations of the diffusing species, a decrease in the transport rate of the diffusing species is reported since the residence time within the trap is longer than for a normal lattice site[38]. We speculate that the presence of traps within tribologically-generated defects accelerates diffusion in a non-parabolic manner.

When diffusion with traps becomes very fast, other mechanisms have to be the rate-determining processes for the oxide growth with varying test duration. For oxidation, different processes are active: adsorption and dissociation of oxygen at the surface and transport and incorporation of oxygen at/in the surface[43]. Surface diffusion of oxygen on copper is known to be much faster than diffusion into the material[43], hence, this is not the rate determining step for the quasi-linear oxide growth. Adsorption and dissociation of oxygen or its incorporation into the surface are thus candidates for determining the oxide growth. However, all three processes may depend on the surface orientation and structure (such as surface steps), temperature, passivation layers and atmosphere (such as oxygen partial pressure)[43–45]. Furthermore, the tribologically-loaded surface is highly complex and constantly evolving under sliding. It is therefore not possible to determine one particular rate-determining factor. Recently, Unutulmazsoy et al. reported a linear rate law for copper oxidation which was explained by the rate-limiting dissociation of oxygen at the oxide–gas surface and thus an interface reaction controlled process[33]. This process is a possible candidate to explain the observed quasi-linear behavior with increasing test duration.

The sliding contact constantly forms and moves dislocations (and grain boundaries) [14,16,46]. Dislocation motion often leaves trapped atoms behind in the lattice while the dislocation moves further. The saturation of trapping sites with diffusing species is thereby avoided and a corresponding decrease in diffusivity is not expected. In the present case, the oxygen left behind in the lattice increases the overall oxygen concentration in the copper and ultimately should lead to a supersaturation of copper with oxygen (the maximum solubility of oxygen in copper is 0.03 at% at 1,340K[47]).

Without this sliding component, the behavior in **Figure 3**c is observed, which basically follows the known growth laws for native oxidation[43]. For static oxidation at room temperature, copper forms a few nanometers thin native oxide layer[48] (SI1). Such a thin layer may also form here after sliding and act as a protective barrier for further oxidation. Oxide growth is thus entirely associated with the tribological loading of the surface. We speculate that defects such as dislocations act as fast diffusion pathways which are kept active during sliding since their motion continuously empties trapping sites. Such fast diffusion channels are needed to explain the observed quasi-linear growth law. We therefore have to pose the question whether there is experimental evidence for dislocation pipe diffusion and segregation of the diffusing species to the pipe in the metal or the oxides.

### 3.3 Experimental evidence for diffusion along dislocations

To answer this question and to identify dislocation pipe diffusion in the complex microstructure of the oxide, we performed atom probe tomography (APT). Specimens were prepared from the middle of the wear track after 5,000 sliding cycles on a high purity copper single crystal with (111) out-of-plane orientation and ±<0-11> sliding direction. **Figure 6**a depicts an oxygen-rich pipe-like feature within the tribologically-deformed copper-subsurface with an oxygen composition of 11 at%. This exceeds the equilibrium solubility of oxygen in copper at room temperature by far. Although APT lacks lattice resolution, similar features were previously attributed to segregation to dislocations by correlation with electron microscopy[49] and can therefore be regarded as a fingerprint of dislocation pipe diffusion[50].

There are two contributions to the oxidation process: (1) oxygen diffusion into the pure metal (**Figure 6**a), which is important in the metal underneath the oxide or as long as no oxide is present; (2) if an oxide is present, oxygen and copper diffuse within the oxide to the interface where the oxide growth is taking place. **Figure 6**b-d present the three-dimensional reconstruction of a section of a hemispherical oxide particle along with iso-surfaces that highlight the metal-oxide interface and two sets of tubular features of different composition inside the oxide but connected to the surface.

The composition profile in **Figure 6**c confirms the nearly 2:1 stoichiometry of Cu and O within the oxide. The metal-oxide interface is slightly depleted in Cu (**Figure 6**c). Interfaces between metals and thermal oxide studied by APT are often found to be somewhat O-rich[51,52]. (A more detailed analysis of the data set from **Figure 6**b is given in SI7.) **Figure 6**d-f show regions of interest that contain Cu-rich pipe-like features within the oxide. The composition profiles along the pipe in **Figure 6**e suggest longitudinal diffusion, with a gradient from Cu-rich at the surface to approx. 65 at% at the metal-oxide interface. Copper and oxygen-rich pathways within the metal or the oxide indicate segregation of copper and oxygen to the defects. Hence, the observed defects serve as effective traps for the diffusing species. When tribological loading is performed in a nitrogen atmosphere followed by exposure to air, no oxide formation was observed (SI6). Thus, the repeated activation of these processes through the shear loading associated with a sliding contact appears to be necessary.

We speculate that the temporal sequence of tribo-oxidation is as follows: first, defects within the metal such as dislocations (**Figure 6**a) or possibly also some grain boundaries accelerate oxidation, ultimately leading to a supersaturation of the copper bulk with oxygen. Second, along with the oxide formation, diffusion pathways within the complex oxides become dominant, such as dislocations or the metal/oxide interface (**Figure 6**b-f). SI5 discusses diffusion along dislocations in metal oxides. Due to the change from metallic copper to an ionic oxide, it is not exactly clear which species diffuses – atoms or ions. It is likely that in the initial stages, copper atoms diffuse which changes with increasing oxidation to copper ions[53],[54].

## 3.4 Late stages of tribo-oxidation

In the late stages of sliding, the oxidation proceeds by diffusion within the oxides and the metal/oxide interface. **Figure 7** depicts samples immediately following tribological deformation for 1,000 cycles at a speed of 1.5 mm/s.

The TEM-image in high angle annular dark field (HAADF) mode from **Figure 7**a shows a layer-like structure after tribological loading. HAADF yields a z-contrast which allows for investigating changes in the chemical composition by a contrast change: darker areas correspond to a lower density or a smaller average atom mass compared to copper. The pores in **Figure 7**a+b are evidence of the Kirkendall effect[55]: For native oxidation, copper ions diffuse faster through the oxide layer than oxygen ions into the bulk[56],[57]. The outward diffusion of copper is compensated by an opposite flux of copper vacancies[53], leading to nearly-spherical pores (d < 50 nm) at the metal-oxide interface (**Figure 7**a). After 48h exposure to ambient air, the metal/oxide interface becomes more even and the pores appear elongated as demonstrated in SI9. The pores are exactly at the position where, as sliding progresses, wear particles delaminate from the sample (**Figure S4**). Such Kirkendall-porosity is associated with diffusion and hence only typically observed at significantly higher relative temperatures[55]. Pores and their reorganization are strong indicators of high defect densities allowing for the vast acceleration of diffusion even under ambient conditions. Results from EDS in the TEM (**Figure 8**) indicate the oxides being (integrally) slightly copper-rich. Previous investigations determined the total oxygen concentration of the oxides to be 23 at % (after 50 cycles), insufficient to form a pure cuprous oxide ($Cu_2O$) with 33 at% of oxygen[12]. All the observations at the late stages of oxidation are in accordance with the model that pipe diffusion along dislocation cores is responsible for the superfast (linear) growth of the oxide under tribological loading. The motion of the dislocations due to tribological loading not only describes the regeneration as fast diffusion paths but also predicts the supersaturation of oxygen within copper and also suggests that a large supersaturation of vacancies in the deformed material allows reorganization processes at ambient conditions thereafter.

The key outcomes of this research are that diffusion along tribologically-induced defects determines the resulting oxides and that the formation and motion of dislocations result in a quasi-linear oxide growth with time. This quasi-linear behavior is limited eventually by thick oxide layers which may spall off (e.g. > 1 µm for 67 h test duration, **Figure 3**b). Deviations from the quasi-linear behavior are also expected for very short test duration at higher speeds (~ 8 minutes for 50 mm/s, **Figure 3**a). The observation of the quasi-linear behavior is important for technical applications: in a sliding contact, larger oxide thicknesses are generated than expected under static conditions, which may lead to early failure of many systems. Understanding the oxide growth and the underlying mechanisms allows for predicting oxide layer thicknesses and remedial actions. For example, reducing the time for out-of-contact diffusion would result in thinner oxide layers or lower oxygen partial pressures would decrease the rate law[33].

**4. Conclusions:**

In conclusion, we have identified that tribologically-induced oxidation is controlled by diffusional processes under mild loading conditions. This is opposite to the currently prevailing, and mostly empirical, understanding of tribo-chemical processes which revolves around a temperature rise at the contact controlling the reaction rate. Such concepts cannot explain our results. By systematically varying the test duration and the exposure time to the ambient environment after tribological loading, we reveal that the oxide growth is almost entirely associated with sliding. As the material in the contact is plastically deformed, dislocation structures and phase boundaries evolve. They are the key diffusion pathways for oxygen to enter the material. An unexpected quasi-linear oxide thickness growth with test duration is explained with the motion of these dislocations which rejuvenates them as diffusion pathways. The ultimate oxide thickness after sliding is most probably determined by the time available for the dissociation of oxygen in-between the passing of the sphere since the trapping of the diffusing species takes place very fast (non-parabolic behavior). The occurrence of Kirkendall-pores at the interface between the oxide and the bulk metal indicates a complex interplay of diffusion fluxes occurring once the oxide has formed. These surprising results open the door to a less empirical and more fundamental understanding of oxidation in frictional contacts, unraveling this important aspect that reduces the lifetime of many mechanical components and systems.

**Acknowledgments:**

We thank P. Schreiber for the support with the 50 mm/s experiments. The authors are grateful to U. Tezins, A. Sturm and C. Bross for their technical support of the atom probe tomography and focused ion beam facilities at the Max-Planck-Institut für Eisenforschung. Funding of this work has partially been provided by the European Research Council under ERC Grant Agreement No. 771237, TriboKey. Additional funding by the German Research Foundation (DFG) under Project No. Gr 4174/1, He 7225/1 and INST 121384/21 is gratefully acknowledged. B.G. is grateful for funding from the ERC-CoG-771602-SHINE.

**Data availability:**

The data that support the findings in this study are available under the link https://doi.org/10.5445/IR/1000117786 and from the corresponding author upon request.

**Appendix A. Supplementary text**

Supplementary text to this article can be found online at:

**Figure 1: Copper oxide formation under tribological loading**. (a) Illustration of the experimental setup: A sapphire sphere sliding against a copper plate. (b) Confocal light microscopy image of the wear track after 1,000 sliding cycles and 1.5 mm/s. Coloring of the surface through oxide formation while sliding. (c) Cross-sectional high angle annular dark field scanning transmission electron microscope image (HAADF-STEM) of the sample in (b). The platinum layer on top of the sample surface was deposited during sample preparation. The atom probe tip represents schematically where tomography samples were cut out with respect to the wear track.

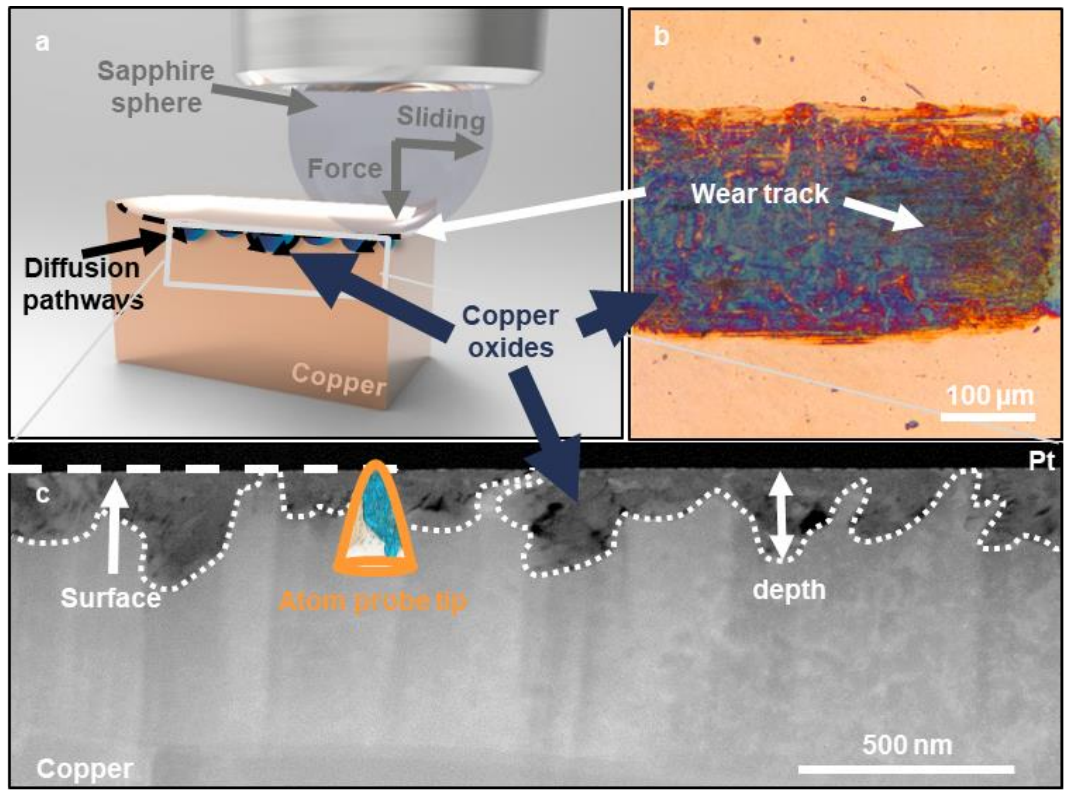

**Figure 2: Oxide thickness as a function of speed and time.** Cross-sectional scanning electron microscopy images (SEM) (a,b, c,d) and scanning transmission electron microscopy (STEM) images (e,f) of the resulting oxides after 1,000 sliding cycles. Variation of sliding speed (a,b), test duration (c,d) and exposure time (e,f). The values in the images represent the most important experimental parameters which were kept constant during the experiment.

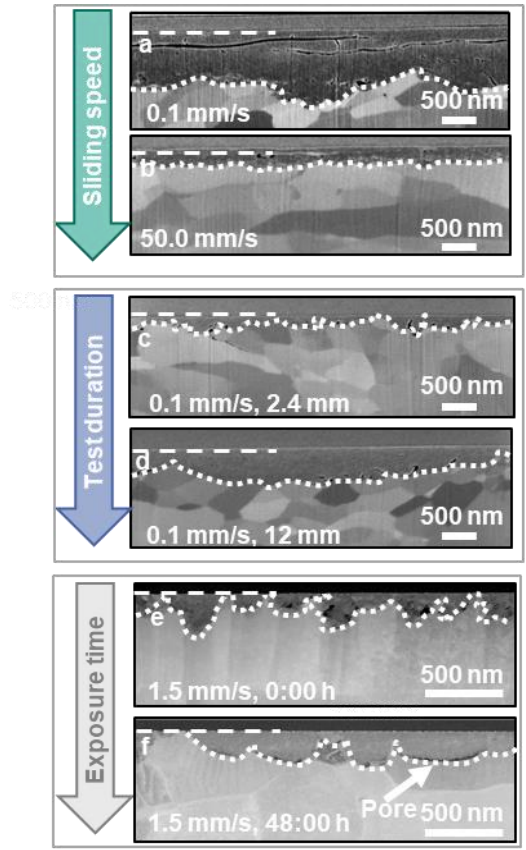

**Figure 3: Quantified oxide thickness** as a function of increasing (a) sliding speed, (b) test duration and (c) exposure time to the ambient environment after the tribological loading of copper polycrystals. To additionally highlight the importance of the test duration, the data presented in (a) for different sliding speeds was also plotted in (b), but this time as a function of test duration. Each data point represents one experiment. The standard deviation stems from the evaluation of multiple SEM images within one cross-section.

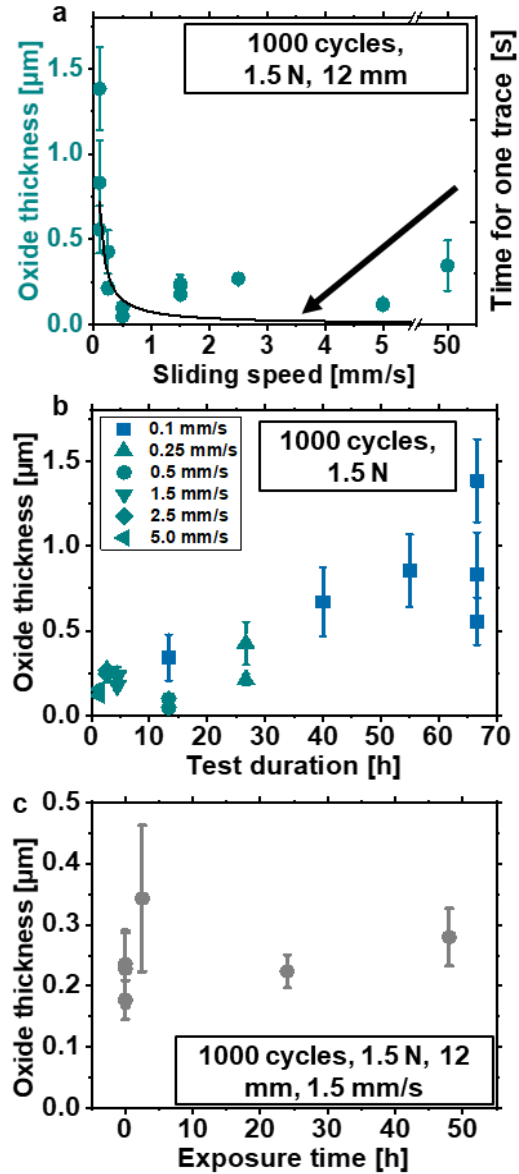

**Figure 4: Confocal light microscopy image of the counter body.** (a) After 1,000 cycles at 1.5 mm/s and (b) 5,000 cycles at 0.5 mm/s.

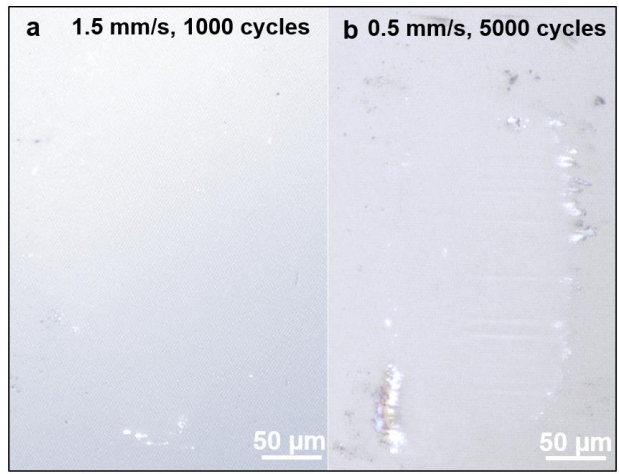

**Figure 5: Diffusion pathways inside oxides through the lattice, along phase boundaries and dislocations.** (a) Side-view and (b) top view. The dimensions represent an oxide after 1,000 sliding cycles with a sliding speed of 0.5 mm/s. $L$ = lattice, $\perp$ = dislocation, PB = phase boundary. $R$ = oxide radius, $t$ = thickness of phase boundary, $a$ = dislocation pipe diameter, $d$ = maximum oxide thickness, $b$ = diffusion path length along phase boundary.

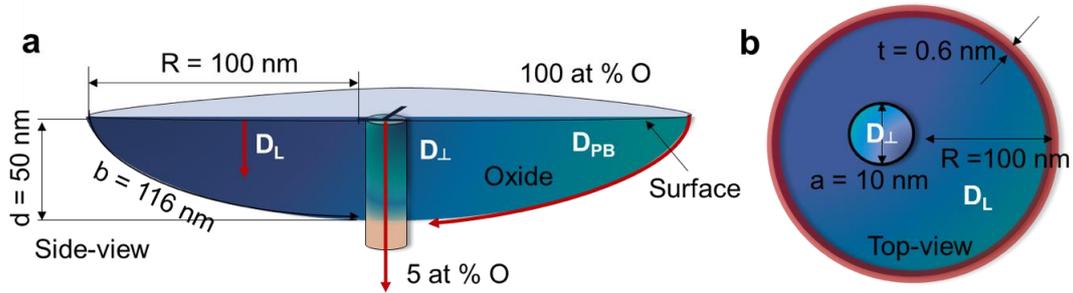

**Figure 6: Diffusion pathways for oxygen into the material.** Atom probe tomography (APT) analysis from the middle of the wear track after 5,000 sliding cycles on a high purity copper single crystal with (111) out-of-plane orientation and ±<0-11> sliding direction. (a) Three-dimensional reconstruction within the tribologically deformed subsurface shows an oxygen pipe feature with 11 at% of oxygen in the Cu matrix. (b) Three-dimensional reconstruction of Cu-oxide particle, with Cu atoms represented as orange dots and oxygen as blue dots. The metal/oxide phase boundary with an iso-composition surface of 12 at% oxygen is depicted. (c) 1-D composition profile across the boundary between the hemispherical particle and copper matrix, the dashed line indicates the average composition in the profile, far from the interface, evidencing Cu depletion at the interface. (d) Three sets of iso-composition surfaces highlight the sharp boundary between the metal and the oxide particle as well as two sets of tubular features (blue = 12 at% O, red = 27 Onm$^{-3}$ and yellow = 32 at% O). (e) Slice through the tomogram showing Cu-rich pipe-like features in the oxide delineated by an iso-point-density surface, which yielded better contrast, as well as (f) one-dimensional composition profiles along the pipe, the dashed line indicates the composition for Cu$_2$O. The evolution of the composition along the pipes is suggestive of diffusional processes.

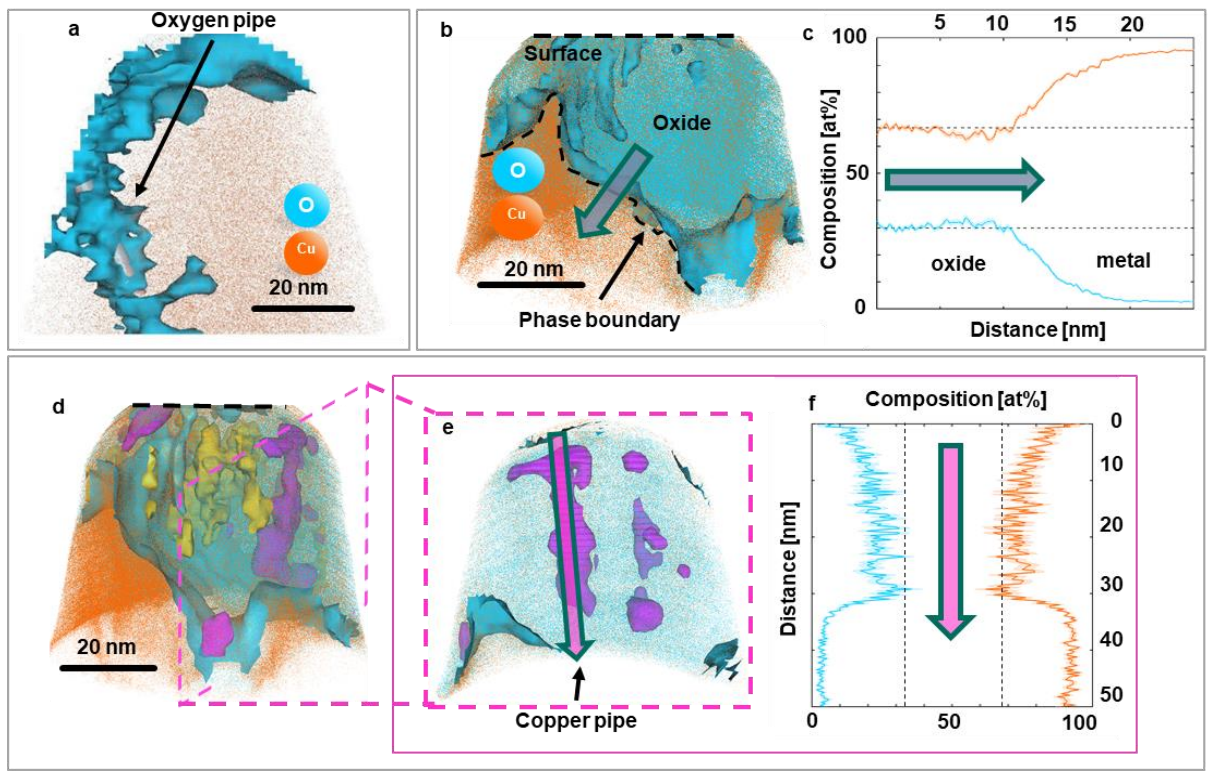

**Figure 7: Structure of oxides immediately following tribological deformation.** (a,b) High-angle annular dark field (HAADF) STEM images of an experiment immediately following tribological deformation with 1.5 mm/s for 1,000 cycles. The white line marks the EDS-line scan depicted in Figure 8.

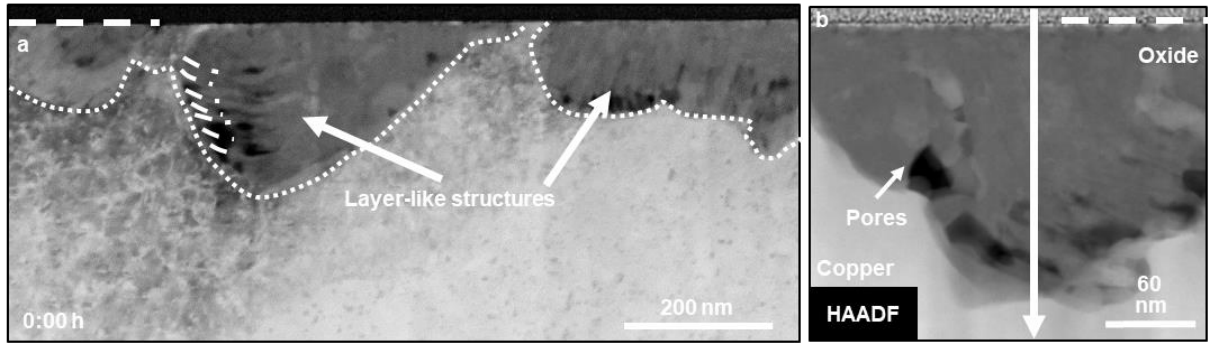

**Figure 8: Chemical composition of oxides immediately following tribological deformation.** Energy-dispersive X-ray spectroscopy (STEM-EDS) measurements of an experiment immediately following tribological deformation with 1.5 mm/s for 1,000 cycles. (a) Copper map of the sample area in Figure 7b; (b) oxygen map of the sample area in Figure 7b. (c) Element-concentration profile along the white line in Figure 7b for copper and oxygen, the dashed lines indicate the composition for $Cu_2O$; The dashed line in represents a $Cu_2O$ composition.

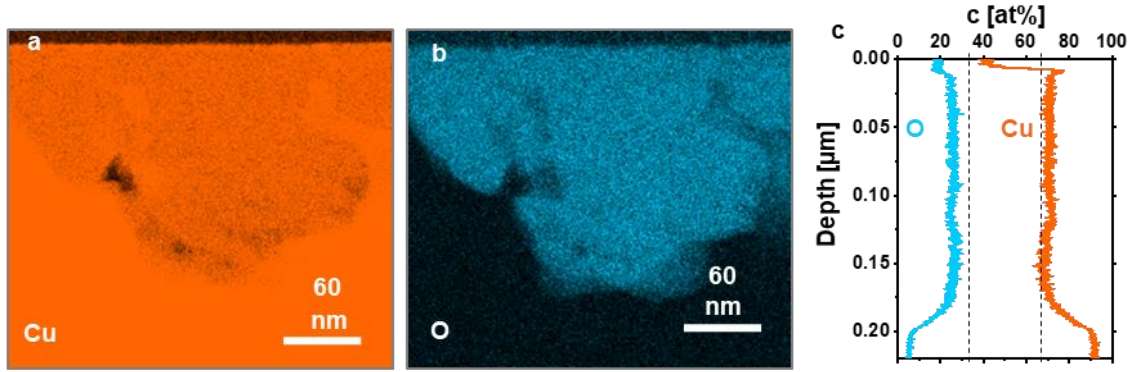

**Table 1: The amount of diffusing atoms per second ($N_x/s$) through the lattice, along phase boundaries and dislocations, calculated with Fick's first law.** $R$ = oxide radius, $t$ = thickness of phase boundary, $a$ = dislocation pipe diameter, $d$ = maximum oxide thickness, $b$ = diffusion path length along phase boundary (see **Figure 5**).

| Diffusion pathways | Lattice | Phase boundary | Dislocation pipe |
|---|---|---|---|
| Area [cm²] | $A_L = \pi R^2$ <br> ~ $3 \cdot 10^{-10}$ | $A_{PB} = \pi(R^2 - (R-t)^2)$ <br> ~ $4 \cdot 10^{-12}$ | $A_{\perp^*} = \pi a^2$ <br> ~ $8 \cdot 10^{-13}$ |
| Diffusion coefficient at RT [cm²/s] | $D_{L^{**}} = 1 \cdot 10^{-34}$ | $D_{BP^{***}} = 1 \cdot 10^{-11}$ | $D_{\perp^{***}} = 1 \cdot 10^{-12}$ |
| Diffusion path length [nm] | d = 50 | b = 116 | d = 50 |
| $\frac{N_x}{s}$ [Number of atoms/s] | $5.1 \cdot 10^{-16}$ | $2.6 \cdot 10^5$ | $1.3 \cdot 10^4$ |

* in good accordance with [58], ** [59], ***for Silver in Silver[60].